\documentstyle[12pt]{article}
\begin{document}

\title{{\huge Strong-Weak Coupling Duality in Quantum Mechanics}}
\author{A. de Souza Dutra\thanks{\noindent Electronic mail:
DUTRA@FEG.UNESP.BR}\\ 
UNESP/Campus de Guaratinguet\'a - DFQ\\
Av. Dr. Ariberto Pereira da Cunha, 333\\
CEP 12500-000\\
Guaratinguet\'a - SP - Brasil}

\date{}
\maketitle
\begin{abstract}
We present a strong-weak coupling duality for quantum mechanical
potentials. Similarly to what happens in quantum field theory, it relates
two problems with inverse couplings, leading to a mapping of the strong
coupling regime into the weak one, giving information from the
nonperturbative region of the parameters space. It can be used to solve
exactly power-type potentials and to extract deep information about the
energy spectra of polynomial ones.
\end{abstract}

\vfill
 \vspace{1.0cm} 
\newpage

Although the solutions of the Schroedinger equation has been studied since
its origin, some of its important features were discovered along the time as
for example the Aharonov-Bohm effect \cite{AB}, the study of quantum
nondemolition measurements \cite{ndm}, coherent states for general
potentials \cite{coherent}, and the geometrical Berry's phase \cite{phase}.
Another interesting feature, is the existence of a number of
analytical solutions for certain polynomial potentials, when some
relations among the potential parameters hold \cite{qes}. Moreover some very
important applications demand for solutions of anharmonic oscillators, as
happens with problems related to tunneling phenomena \cite{tunnel}.
Furthermore they can also be used in order to study models analogous to some
quantum field theory ones \cite{Martin}.

In this letter we present a very simple and, at same time, very powerful
duality for potentials in quantum mechanics. It is capable of
defining the analytical dependence of the energy spectrum on the potential
parameters, for a large class of potentials. In fact it is just a matter of
experimentation to confirm the validity of this feature for potentials like
the harmonic oscillator, the Coulomb potential and other exactly solvable
potentials. The principal goal of this letter is to show that potentials
which can not be treated through perturbation theory are connected with
others which can, letting one to use this connection to get the energy in an
strong coupling regime through perturbative calculations in the dual
potential. This is very similar to the recently studied electric-magnetic
duality in $N=2$ supersymmetric Yang-Mills theory \cite{witten}. In
addition we get the exact behavior of the energy spectra of the 
general power potential $V(x)=ax^N$ \cite{Arponen}. We also discuss its
application in polynomial potentials \cite{poli}, \cite{Leacock}, \cite
{tonico}.

So let us begin by setting the problem in a general way. We start with the
one-dimensional time-independent Schroedinger equation

\begin{equation}
-\frac{\hbar ^2}{2\mu }\frac{d^2\psi (x)}{dx^2}+V(x)\psi (x)=E\psi (x), 
\end{equation}

{\noindent}perform a dilatation in the spatial coordinate $x\rightarrow gx$,
so that

\begin{equation}
-\frac{\hbar ^2}{2\mu }\frac{d^2\psi (x)}{dx^2}+U(x)\psi (x)={\cal E}\psi
(x), 
\end{equation}

{\noindent}with $U(x)\equiv g^2V(gx)$ and ${\cal E} \equiv g^2E$.

As an example we consider the polynomial potential $V(x)=a\;x^6+b\;x^4+c\;x^2
$ \cite{poli}. In this case one has:

\begin{equation}
U(x)\;=\;\tilde a\;x^6+\tilde b\;x^4+\tilde c\;x^2,\;\;\left( \tilde a\equiv
a\;g^8,\;\tilde b\equiv b\;g^6,\;\tilde c\equiv c\;g^4\right) , 
\end{equation}

\noindent so that, if $a$ is great compared with $c$, it is not possible to
apply the usual perturbation theory, and even if one use the asymptotic
approach like that in \cite{poli}, valid for great values of the principal
quantum number, the problem for the low values of $n$ is still maintained.
If however one chooses a conveniently small $g$, it is possible to invert
the situation, obtaining a perturbative potential where $\tilde c$ is
much greater than $\tilde a$. In fact, for a given set of potential
parameters, a family of potentials have their energy fixed due to this
symmetry. As an example one can compare two numerical results: ($%
a=100,b=1,c=1$) with $E_0\,=\,13.042\,036\,330$, and ($%
a=10^{-6},b=10^{-6},c=10^{-4}$), obtained from the first setting $g=10^{-1}$%
, with $E_0\,=\,0.130\,420\,363$, which is in accordance with the respective
numerical result. Other important situations can be treated in a quite
similar way, as for instance the problem of atoms in strong magnetic fields 
\cite{rydberg}, interaction of anyons \cite{anyon}, etc.. . In all these
cases, one can make a connection between the weak and the strong coupling
regime.

Let us now see that this quite simple idea is almost powerful enough to
determine by itself the energy spectra for the power type potentials. In
this case $V(x)=ax^N$, and after making the above transformations and
imposing the covariance of the energy, one gets

\begin{equation}
E_n=k_n^{(N)}\left( \frac{\hbar ^2}\mu \right) ^{\frac N{N+2}}a^{\frac
2{2+N}}, 
\end{equation}

{\noindent}where $k_n^{(N)}$ is a factor which can not be determined by
the duality, and that for $N=2$ (harmonic oscillator) will be equal
to $n+\frac 12$. For the other potentials it is only necessary determine it
for a single value of $a$ ($a=1$ for example) and the remaining will stay
fixed. One can take for instance the cases with $N=4$ and $N=6$. In these
cases the first five levels are such that one get the results in table I.

Now if one wishes to obtain the value of the energy of any level, with other
values of $a$ and $\mu $ (and using the correct value of $\hbar $) is just a
question of substitution and evaluation of Eq.(4), using the constants
obtained in table I. Furthermore one can also obtain exactly the wave
function by solving the Schroedinger equation, getting an infinite series
depending on the energy. For $N=6$, one has

\begin{equation}
\Psi (x)\;\equiv \;\sum_{n=0}^\infty \;a_nx^n\;e^{-\frac{\gamma x^4}4} 
\end{equation}

\noindent where $\gamma \equiv \sqrt{2a}$ and the coefficients $a_n$ should
be computed through the recurrence equation $a_{n+2}=\frac{-\left[
2Ea_n-\left( 2n-1\right) \gamma \;a_{n-2}\right] }{\left( n+1\right) \left(
n+2\right) }$. So, substituting the exact energy obtained above, one gets
the corresponding exact wave function.

One can do a comparison between the energy obtained with the method
developed above and that calculated through numerical calculations for some
different values of the potential parameter $a$ where, for sake of
simplicity, we used $\hbar =\mu =1$. In doing so one obtain the figures
appearing in table II. Note that the only difference, when it exists, is in
the last decimal due to the necessary approximation. Nevertheless, if one
calculates with even more precision one sees that this is only a consequence
of the approximation.

As our next application of this duality, we consider a polynomial potential
like $V(x)=a\;x^N+b\;x^M+c\;x^L$ ($M$ and $L$ lesser than $N$). As for
large values of $x$, the $N$-power term dominates, the energy will have its
dependence factored and the dependence on the remaining parameters is such
that one can write down a series invariant under this symmetry. So we will
have something like

\begin{equation}
E_n=a^{\frac 2{2+N}}\sum_{m=0}^\infty b_mx^m, 
\end{equation}

\noindent where $b_0=k_n^{(N)}\left( \frac{\hbar ^2}\mu \right) ^{\frac
N{N+2}}$, and $x$ must be invariant under this symmetry, so we conclude that 
$x=a^\alpha b^\beta c^\gamma $, with the coefficients $\alpha ,\beta ,\gamma 
$ obeying the equation $\alpha \left( N+2\right) +\beta \left( M+2\right)
+\gamma \left( L+2\right) =0$, in order to keep the symmetry. In fact one
can easily verify that the polynomial potentials treated in \cite{poli} and 
\cite{tonico} have a series where the corresponding expanding variable obeys
this equation, so preserving the symmetry outlined here. However in both
cases the series diverges when $a$ vanishes, but this feature can be
an artifact of the series expansion. One could for example in the
potential with $N=6,\, M=2\, L=0$, try two
non-singular functions: $E_n=\left[ \alpha _n^2\,a^{\frac 12}+\beta
_n^2\,b\right] ^{\frac 12}$ or $E_n=\left[ \alpha _n^4\,a+\beta
_n^4\,b^2\right] ^{\frac 14}$, which have as expansion series, at a given
order, singular expressions like that in (6). Furthermore, it can be
verified that they represent upper and lower approximations for the exact
levels respectively, with errors below ten percent. Now we are looking
for more precise analytically approximated energies.

\vspace{1.0cm}

\noindent ACKNOWLEDGEMENTS: This work was partially supported by CNPq,
FUNDUNESP and FAPESP (contract \#95/4795-8). The author is also indebted
to A. S. de Castro for helpfull discussions.

\newpage

\noindent TABLE I: $k_n^N$ coefficients for $N$ equal to 4 and 6, for the
first five energy levels. \vspace{0.5cm}

\begin{center}
\begin{tabular}{|c|c|c|} \hline
\multicolumn{3}{|c|}{TABLE I} \\ \hline
n & $k_n^{(4)}$ & $k_n^{(6)}$ \\ \hline
0 & 0.66798625 & 0.68070361 \\ \hline
1 & 2.39038996 & 2.56845109 \\ \hline
2 & 4.69361420 & 5.38898250 \\ \hline
3 & 7.34950780 & 8.88132184 \\ \hline
4 & 10.24505604 & 12.80383861 \\ \hline
\end{tabular}
\end{center}

\vspace{0.5cm}

\noindent TABLE II: Comparison among the energies obtained from equation (4)
and from numerical calculations for $N\,=\,6$, $a=2$ and $a=\pi$. 
\vspace{0.5cm}

\begin{center}
\begin{tabular}{|c|c|c|c|c|} \hline
\multicolumn{5}{|c|}{TABLE II} \\ \hline
n & $E_n(a=2)$ & $E_{numerical}$ & $E_n(a=\pi )$ & $E_{numerical}$ \\ \hline
0 & 0.80949758  & 0.80949758  & 0.90624478 & 0.90624478   \\ \hline
1 & 3.05442031 & 3.05442030 & 3.41946977 & 3.41946976  \\ \hline
2 & 6.40861633 & 6.408621632 & 7.17454298 & 7.17454297  \\ \hline
3 & 10.56173112 & 10.56173111 & 11.82401784 & 11.82401785  \\ \hline
4 & 15.22641597 & 15.22641598 & 17.04620314 & 17.04620315  \\ \hline
\end{tabular}
\end{center}

\end{document}